\documentstyle[aps,twocolumn,pra,graphicx]{revtex}
\newcommand{\smallperp}{{\scriptscriptstyle \perp}}
\begin{document}
\title{Two-component Bose-Einstein Condensates with Large Number of
  Vortices} \author{Erich J.~Mueller\cite{EJM} and Tin-Lun
  Ho\cite{TLH}} \address{Department of Physics, The Ohio State
  University, Columbus, Ohio 43210}
\date{Revised: March 13, 2002}

\wideabs{ 
\maketitle

\begin{abstract}
We consider the condensate wavefunction of a rapidly rotating
two-component Bose gas with an equal number of particles in each
component.
If the interactions % constants 
between like and unlike species are very similar (as occurs for two
hyperfine states of $^{87}$Rb or $^{23}$Na) we find that the two
components contain identical rectangular vortex lattices, where the
unit cell has an aspect ratio of $\sqrt{3}$, and one lattice is
displaced to the center of the unit cell of the other.  Our results
are based on an exact evaluation of the vortex lattice energy in the
large angular momentum (or quantum Hall) regime.
\end{abstract}

}

Experiments on rotating Bose gases have progressed rapidly in the last
two years. Soon after the pioneer work at JILA \cite{JILA1} and ENS
\cite{Paris}, the MIT group created a vortex lattice with as many as
160 vortices\cite{MIT}.  Recently, the JILA group has invented an
ingenious method to increase the angular momentum of a condensate by
performing evaporative cooling on a rotating normal cloud\cite{JILA}.
In this process, the system spins faster and faster as it is cooled,
while remaining close to equilibrium. With such rapid progress, one
expects that equilibrium Bose gases with even larger angular momenta
may be produced in the near future.

At present, most experiments on vortex lattices are performed in
single component Bose systems. It is natural to ask what happens in
two-component Bose gases, such as those made up of two hyperfine spin
states of the same atom. The vortex lattices in such systems are bound
to be more intricate than those in single component condensates, as
the vortices in different components can move relative to one another
to minimize the energy. The purpose of this paper is to study the
vortex lattices of two-component systems with large number of
vortices, in what we call the ``mean-field quantum Hall regime''.
This is the regime where mean field theory remains valid so that each
component (labeled by an index ``$i$'', $i=1,2$) is characterized by a
condensate wavefunction $\Psi_{i}$; yet the angular momentum of the
system is so high that $\Psi_{i}$ is made up of the orbitals in the
lowest Landau level in the plane perpendicular to the rotation axis.
It has been shown recently\cite{Ho} that this regime will emerge in a
{\em three dimensional} Bose gas at sufficiently high angular momenta
\cite{horeq}.
We focus on this regime because the wavefunction in this limit
acquires an analytic structure which allows exact evaluation of the
energy of a vortex lattice. As a result, it is possible to scan
through a wide range of lattice structures which would be impractical
for numerical calculations because of the time and the accuracy
required.  Although not directly applicable to current experiments on
vortex lattices (which are performed at lower angular momenta), the
physics of the mean field quantum Hall regime is still quite relevant
as the vortex lattices in these two regimes are connected continuously
to each other.

One special feature of the majority of two-component gases so far
studied, (notably mixtures of hyperfine states of $^{87}$Rb
\cite{JILAtwospecies} in magnetic traps, or $^{23}$Na
\cite{MITtwospecies} in optical traps) is that the interactions between
like species (denoted $g_{1}$ and $g_{2}$) and unlike species (denoted
$g_{12}$) are very similar, within a few percent of each other. Thus,
if there are an equal number of bosons in each component, and if the
trapping potentials of the two components are made identical, 
then the two components will be the same size and contain the same
density of vortices.  In
this case, one expects that each component will contain identical
vortex lattices, with one lattice displaced relative to the other.
While we are mainly interested in the experimentally relevant cases,
where $g_{1}\sim g_{2}\sim g_{12}$, considerable insight is gained by
studying vortex lattices as a function of the interactions.
Considering the case $g_{1}\sim g_{2} \neq g_{12}$, we find a wide
range of vortex lattice structures as the parameter $\alpha =
g_{12}/\sqrt{g_{1}g_{2}}$ is varied.  The vortex lattice has a fixed
structure over certain intervals of $\alpha$, while it varies
continuously in others. The structure near the isotropic point
$g_{1}=g_{2}=g_{12}$ consists of identical rectangular lattices in
both components, with one displaced to the center of the unit cell of
the other. The aspect ratio of the unit cell changes with $\alpha$,
and is $\sqrt{3}$ when $\alpha=1$.

{\bf The mean-field quantum Hall regime:} The condensate wavefunctions
$\Psi_{1}$ and $\Psi_{2}$ of a two-component rotating Bose gas are
determined by minimizing the grand potential $K = E -\Omega L_{z} -
\mu_{1} N_{1} -\mu_{2}N_{2}$, where $E$ is the energy of the system,
$\Omega$ is the rotational frequency, $L_{z}$ is the angular momentum
along $z$, and $\mu_{i}$ ($i=1,2$) are the chemical potentials fixing
the number of bosons $N_{1}$ and $N_{2}$ in each component.  For
simplicity, we assume identical trapping potentials for each
component. We consider a cigar shaped trap with the symmetry axis $z$
coinciding with the axis of rotation.  As discuss in \cite{Ho}, the
slow variation of the trapping potential along $z$ allows one to apply
a Thomas-Fermi approximation for the $z$ dependence of $\Psi_{i}$ and
write $K$ as $\int\!{\rm d}z{\rm d}{\bf r}\, {\cal K}({\bf r}, z)$,
\begin{eqnarray}\label{genk}
&{\cal K}({\bf r}, z)  = \sum_{i=1,2} \Psi^{\ast}_{i}
[ h - \mu_{i}(z)]\Psi^{}_{i} + {\cal V},&\\
&h = \frac{1}{2M}\left(\frac{\hbar}{i} {\bf \nabla} - M\Omega
\hat{\bf z} \times {\bf r}\right)^2
+  \frac{1}{2}M(\omega^{2}_{\smallperp}- \Omega^2)r^2,&\\
&{\cal V} = \frac{1}{2}g_{1}|\Psi_{1}|^{4} + \frac{1}{2}g_{2}
|\Psi_{2}|^{4} + g_{12}|\Psi_{1}|^2|\Psi_{2}|^2,&
\end{eqnarray}
with ${\bf r}=(x,y)$,
$\mu_{i}(z)=\mu_{i}-\frac{1}{2}M\omega_{z}^{2}z^2$, $g_{i}=
4\pi\hbar^2 a_{i}/M$, $(i=1,2)$, and $g_{12}= 4\pi\hbar^2 a_{12}/M$,
where $a_{i}$, and $a_{ij}$ are the s-wave scattering lengths between
like and unlike bosons respectively. As $z$ is treated as a parameter,
it is convenient to write $\Psi_{i}=\sqrt{n_{i}(z)}\Phi_{i}({\bf r};
z)$, with $\int |\Phi_{i}({\bf r}; z)|^2\, d^{2}r =1$.  The number
constraint $\int\!{\rm d}{\bf r}{\rm d}z\,|\Psi|^2 = N_{i}$ becomes
$\int\! n_{i}(z)\,{\rm d}z = N_{i}$.

As $\Omega$ approaches $\omega_{\smallperp}$, the wavefunctions
$\Phi_{i}$ are made up of the orbitals $u_{m}({\bf r})$ in the lowest
Landau level in the $xy$-plane, $\Phi_{i}({\bf r},z) =
\sum_{m=0}^{\infty} c_{m}(z)u_{m}({\bf r})$, where $u_{m}({\bf r}) =
(2\pi m!)^{-1/2} [(x+iy)/d]^{m}e^{-r^2/2d^2}$, and
$d=\sqrt{\hbar/M\omega_{\smallperp}}$.  The potential ${\cal K}$ then
becomes
\begin{equation}
{\cal K} = \sum_{i=1,2} \left[ \hbar(\omega_{\smallperp}\!-\Omega)
\frac{\langle r^2 \rangle_{i} }{d^2} -  
\mu_i(z)+\hbar\omega_{\smallperp}
\right]
 n^{}_{i}(z) + {\cal V}
\label{K} \end{equation}
where $\langle r^2 \rangle_{i} = \int\! r^2 |\Phi_{i}|^2 \, d^2 r$,
and
\begin{equation}
{\cal V} = \int\! {\rm d}^2 r \left[ \frac{1}{2} \sum_{i=1,2}
g_{i} n^{2}_{i}
 |\Phi_{i}|^4  + g_{12} n^{}_{1} n^{}_{2}
|\Phi_{1}|^2 |\Phi_{2}|^2  \right].
\end{equation}
As shown in ref.\cite{Ho}, wavefunctions in the lowest Landau level
(not normalized) can be written as
\begin{equation}\label{sigmatheta}
\phi({\bf r}) = \lambda \prod_{\alpha}(w - a_{\alpha})
e^{- r^2/2d^2}, \,\,\,\,\,\, w = x + iy.
\end{equation}
where $\lambda$ is an arbitrary constant and $\{ a^{}_{\alpha} \}$ are
the zeros of $\phi$. If the zeros form a infinite lattice with unit
cell size $v_{c}$, it is shown in \cite{Ho} that $|\phi|^2$ is a
product of a Gaussian and a function periodic under lattice
translation. i.e.
\begin{equation}
|\phi|^2 =  e^{-r^2/\sigma^2} g({\bf r}),  \,\,\,\,\,\,
g({\bf r}) = g({\bf r} + {\bf R}),
\label{new} \end{equation}
where ${\bf R} = n_{1}{\bf B}_{1} + n_{2}{\bf B}_{2}$, $n_{i}$ are
integers, and ${\bf B}_{1}$, ${\bf B}_{2}$ are the basis vectors of
the lattice. The width $\sigma$ reflects the number of vortices of the
system. It is given by
\begin{equation}
 \sigma^{-2} = d^{-2} - \pi v_{c}^{-1}.
\label{sigma} \end{equation}
The periodicity of $g({\bf r})$ implies $g({\bf r}) =
v_{c}^{-1}\sum_{\bf K} g^{}_{\bf K} e^{i{\bf K}\cdot {\bf r}}$, where
$\{ {\bf K} \}$ are the reciprocal lattice vectors.

In the following, we shall consider a two-component Bose gas with
equal particle numbers and trapping potentials, and with interactions
$g_{1}\sim g_{2} \neq g_{12}$. If $g_{1}=g_{2}$, the two components
are identical and one expects that 
each will contain identical vortex lattices,
translated with respect to one another.  Sufficiently small
differences in $g_{1}-g_{2}$ should not change this structure (though
changes may occur in the density profiles $n_i(z)$, the parameters of the
lattice, and the relative
displacement $r_0$).  This structure persists 
because, even when $g_1\neq g_2$, the two
components contain equal vorticity, hence an equal
density of vortices.  The potential energy is minimized by
interlacing the two lattices; if the vortex
lattice in one component were to deform, the other has to follow to
keep the interaction energy at a minimum.  
We shall therefore consider {\em normalized} condensates
$\Phi_{1}$ and $\Phi_{2}$ in Eq. (\ref{K}) with densities of the form
\begin{eqnarray}
|\Phi_{1}|^2 &=& (\pi \sigma^2)^{-1} \textstyle \sum_{{\bf K}} \tilde{g}_{\bf K}
e^{i{\bf K}\cdot {\bf r}} e^{-r^2/\sigma^2}
\label{Phi2}\\
|\Phi_{2}|^2 &=& (\pi \sigma^2)^{-1}\textstyle\sum_{{\bf K}} \tilde{g}_{\bf K}
e^{i{\bf K}\cdot ({\bf r}-{\bf r}_{o})} e^{-r^2/\sigma^2}
\label{Phi4}\\
\tilde{g}_{\bf K} &=& {g}_{\bf K} / \left(\textstyle\sum_{{\bf K'}} g_{\bf K'} e^{-
\sigma^2 {\bf K'}^2/4}\right).
\label{gK} \end{eqnarray}
The wavefunction is described by variational parameters $n_{i}(z)$,
$\sigma^2$, the basis vectors ${\bf B}_{i}$ (which determine the unit
cell size $v_{c}$), and the relative displacement ${\bf r}_{o}$.

By integrating Eqs.~(\ref{Phi2}) and (\ref{Phi4}), one sees that up to
terms of relative order $v_c/\sigma^2$ the cloud's radius is $\langle
r^{2} \rangle_{1}= \langle r^{2} \rangle_{2}= \sigma^2$.  Defining the
quantities $I$ and $I_{12}$ as $\int\! |\Phi_{i}|^4\, {\rm d}^2 r
\equiv I/ (\pi \sigma^2)$ and $\int\! |\Phi_{1}|^2 |\Phi_{2}|^2\,{\rm
  d}^2 r = I_{12}/(\pi\sigma^2)$, we have
\begin{eqnarray}
I &=& \sum_{{\bf
K}, {\bf K'}} \tilde{g}^{}_{\bf K} \tilde{g}^{}_{\bf K'} e^{-\sigma^2
|{\bf K} + {\bf K'}|^2/4}  ,
\label{I}\\
I_{12} &=&
\sum_{\bf K}\tilde{g}^{}_{\bf K}  \tilde{g}^{}_{\bf K'}  e^{-i{\bf K'}\cdot
{\bf r}_{o}} e^{-\sigma^2 |{\bf K} + {\bf K'}|^2/4},
\label{I12} \end{eqnarray}
and the potential ${\cal K}$ takes the form
\begin{eqnarray}\label{kform}
{\cal K} =& -\left[\mu(z) -\hbar\omega_{\smallperp} -
\hbar(\omega_{\smallperp}-\Omega) (\sigma^2/d^2)  \right](n_{1}+n_{2})
\nonumber \\
 & + (d^2/2\sigma^{2}) \left( n_{1}^2 g_{1}I +  n_{2}^2 g_{2}I +
2n _{1}n_{2}g_{12}I_{12} \right).
\label{newK} \end{eqnarray}
To evaluate ${\cal K}$, we need an explicit expression for the
coefficients $g_{\bf K}$.  These coefficients are most easily derived
by introducing the complex representation for the basis vectors,
$b_{i} \equiv (\hat{\bf x} + i \hat{\bf y})\cdot {\bf B}_{i}$. The
area of the unit cell is then $v_{c}^{} = i(b_{1}^{\ast} b_{2} -
b_{2}^{\ast} b_{1})/2$.  If we orient the lattice so that $b_{1}$ is
real, i.e. ${\bf B}_{1} = b_{1}\hat{\bf x}$, ${\bf B}_{2}=
b_{1}(u\hat{\bf x} + v\hat{\bf y})$, we then have
\begin{equation}
b_{2} \equiv  b_{1}(u + iv), \,\,\,\,\,\,\, v_{c}^{} = b_{1}^{2} v.
\label{uv} \end{equation}
In the Appendix, we show that {\em a function $\Phi$ in the lowest
  Landau level describing a regular vortex lattice contained in a
  cylindrically symmetric cloud will have the form $\phi({\bf r}) =
  f(w)e^{-r^2/2d^2}$, with $w\equiv x+iy$, and
\begin{equation}
f(w) = \theta \left(\zeta, \tau \right) e^{\pi
w^2/2v_{c}}
\label{f} \end{equation}
where $\zeta = w/b_1= (x+iy)/b_1$, $\tau=u+iv=b_2/b_1$, 
and $\theta$ is the Jacobi theta-function
defined as
\begin{equation}\label{theta}
\theta(\zeta, \tau) = \frac{1}{i} \sum_{n= -\infty}^{\infty}
(-1)^{n} e^{i\pi \tau (n + 1/2)^2} e^{2\pi i \zeta(n+1/2)}.
\end{equation}
The density $|\phi|^2$ is therefore of the form Eq. (\ref{new}), with
$\sigma$ given in Eq. (\ref{sigma}), and }
\begin{equation}
g({\bf r}) = |\theta \left(\zeta, \tau \right){\em exp}(-\pi
y^2/v_{c})|^2.
\label{key} 
\end{equation}
The Jacobi theta-function has the quasi-periodic properties
\begin{eqnarray}
\theta(\zeta +1 , \tau) &=& \theta(\zeta, \tau),
\\
\theta(\zeta + \tau, \tau)
&=& - e^{-i\pi(\tau + 2\zeta)}\theta(\zeta, \tau),
\end{eqnarray}
which implies the periodic property $g({\bf r})= g({\bf r}+{\bf R})$.
The Fourier coefficients of $g({\bf r})$ are
\begin{equation}
g^{}_{\bf K} = (-1)^{m_{1}+m_{2} + m_{1}m_{2}} 
e^{-v^{}_{c}|{\bf K}|^2/8\pi} \sqrt{\frac{v^{}_{c}}{2}} \,\,\, , 
\label{gK2} \end{equation}
where ${\bf K}=m_{1}{\bf K}_{1}+ m_{2}{\bf K}_{2}$, and ${\bf K}_{i}$
are the basis vector of the reciprocal lattice, ${\bf K}_{1} =
(2\pi/v_{c}){\bf B}_{2}\times \hat{\bf z}$, ${\bf K}_{2} =
(2\pi/v_{c})\hat{\bf z} \times {\bf B}_{1}$, and
\begin{equation}
v_{c} {\bf K}^2 = \frac{(2\pi)^2}{v}\left( (vm_{1})^2 + (m_{2} -
um_{1})^2\right)
\end{equation}
Since we work in the limit of large vortex number, the size of the
cloud is much larger than the unit cell, i.e. $\pi \sigma^{2}/v_{c}
>>1$. We can therefore ignore all ${\bf K}+{\bf K'}\neq 0$ terms in
Eqs. (\ref{I}) and (\ref{I12}), since $\sigma^2 {\bf K}^2 > \pi
\sigma^2/v_{c}$. We then have
\begin{equation}
I = \sum_{\bf K} \left|\frac{g_{\bf K}}{g_{\bf 0}}\right|^2,
\,\,\,\,\,
\,
I_{12} = \sum_{\bf K} \left|\frac{g_{\bf K}}{g_{\bf 0}}\right|^2
{\rm cos}{\bf K}\cdot {\bf r}_{o},
\label{newI} \end{equation}
where $g_{\bf K}^{}$ is given by Eq. (\ref{gK2}) and ${\bf K}$-sum is
over the integers $m_{1}, m_{2}$.  Since the expressions of $I$ and
$I_{12}$ in Eq. (\ref{newI}) are independent of $\sigma^2$, the
minimization of ${\cal K}$ in Eq. (\ref{newK}) with respect to
$\sigma^2$ and $n_{i}$ become very simple. The optimum $\sigma^2,
{\bf n}=(n_1,n_2)$, and ${\cal K}$ are given by
\begin{eqnarray}\label{opt1}
\sigma^2 &=& d^2\left( \mu(z)-\hbar\omega_\smallperp\right)/
\left( 3\hbar (\omega_{\smallperp} - \Omega)\right),
\\
%{\bf n}(z)
%&=& \frac{2(\mu(z) - \hbar\omega_{\smallperp})^2}
%{9\hbar (\omega_{\smallperp} -
%\Omega)} {\bf G}^{-1}
%\left( \begin{array}{c} 1\\ 1 \end{array}\right),
{\bf n}(z) &=& 
(2/3)(\sigma^2/d^2) (\mu(z)-\hbar\omega_\smallperp)\,\,
{\bf G}^{-1}\cdot{\bf 1},
%\textstyle
%\left( \textstyle\small \begin{array}{c}\tiny 1\\
%\Huge 1 \end{array}\right),
\label{result1} \\
%{\cal K} &=&  - \frac{2}{27}
%\frac{(\mu(z)-\hbar\omega_\smallperp)^3}{\hbar(\omega_\smallperp-\Omega)}
%(1 1)\cdot {\bf G}^{-1}\cdot \left( \begin{array}{c} 1\\ 1
%\end{array}\right),
{\cal K} &=& - (1/3) (\mu(z)-\hbar\omega_\smallperp)\,\,
%\left(\begin{array}{cc}1&1\end{array}\right)
{\bf 1}\cdot {\bf n}(z),
\\
{\bf G} &=& \left(
\begin{array}{cc} g_{1} I & g_{12}I_{12}\\ g_{12}I_{12} & g_{2} I
\end{array}\right),\,\,\,\,\,
{\bf 1} =
\left(
\begin{array}{c}
1\\
1\\
\end{array}
\right),
\label{result2} \end{eqnarray}
It is clear from Eqs. (\ref{opt1}) through (\ref{result2}) that the
solution for the case where $g_{1}-g_{2}<<|g_{1}+g_{2}|$ is very close
to that of $g_{1}=g_{2}$.  
The lattice shape (parameterized by ${\bf r}_0$, $u$,
and $v$) enters the grand potential only
through the factor ${\bf 1\cdot G^{-1}\cdot 1}$.  When $g_1=g_2$ this
term is inversely proportional to
$J=I+\alpha I_{12}$,  and the most favorable lattice
is the one that minimizes $J$.

{\bf Summary of Results:} It is interesting to compare the
two-component case with the single-component case. In the latter system,
energy minimization reduces
to minimizing $I$.  The only local minimum is the triangular lattice,
where $I=1.1596$; the square lattice is a saddle point with
$I=1.1803$.  The minute difference between these values of $I$ makes a
simple numerical minimization of (\ref{genk}) challenging and
illustrates the utility of the analytic scheme used here.

For a two-component Bose gas, the most favorable lattice minimizes
%with $g_{1}=g_{2}=g$, miminization of
%${\cal K}$ in Eq. (\ref{result2}) with respect to ${\bf r}_{o}$, $u$,
%and $v$ reduces to miminizing the quantity 
$I+\alpha I_{12}$.  In the
minimization it is convenient to measure lengths in unit of the the
basis vector ${\bf B}_{1} = b_{1}\hat{\bf x}$, and write complex
representation of ${\bf B}_{2}$ and ${\bf r}_{o}$ as $\tau=
u+iv=|\tau|e^{i\eta}$ and $r_{o}\equiv a+b\tau$ respectively.  The
phase diagram of the vortex lattice as a function of the ratio
$\alpha= g_{12}/g$ is shown in Figs. 1 and 2. The major features are

\noindent (a) $\alpha <0$: In this region the vortices of the two components 
coincide with each other ($a=b=0$) to form a triangular lattice
($\tau=e^{i \pi/3}$).

\noindent (b) 
$0<\alpha< 0.172$: the vortex lattice in each component remains
triangular.  However, $r_{o}$ undergoes a first order change so that
one lattice is displaced to the center of the triangle of the other
($a=b=1/3$).  The lattice type (characterized by $\tau=e^{i\pi/3}$)
remains constant within this interval.

\noindent 
(c) $0.172< \alpha= 0.373$: At $\alpha= 0.172$, ${\bf r}_{o}$ jumps
from the center of the triangle (i.e. half of the unit cell) to the
center of the (rhombic) unit cell ($a=b=1/2$).  The angle $\eta$ jumps
from $60^{o}$ to $67.95^{o}$ at $\alpha= 0.172$, and increases
continuously to $90^{o}$ as $\alpha$ increases to $0.372$. As a
result, the lattice type is no longer fixed and the unit cell is a
rhombus.  The modulus of $\tau$, however, remains fixed across this
region.

\noindent 
(d) $0.373<\alpha<0.926$: The two lattices are ``mode-locked'' into a
centered square structure throughout the entire interval
$(\tau=i,a=b=1/2)$.

\noindent (e) $0.926<\alpha$: The
lattice type again varies continuously with interaction $\alpha$.
Each component's vortex lattice has a rectangular unit cell
($\eta=\pi/2$) whose aspect ratio $|\tau|$ increases with $\alpha$.
Both $^{87}$Rb and $^{23}$Na have interaction parameters within this
range.  At $\alpha=1$, ($g_{1}=g_{2}=g_{12}=g$), the aspect ratio is
$\sqrt{3}$.  If one ignores the difference between the components, the
combined lattice is triangular, as is expected.

It is interesting to note that in the absence of rotation, the two
components change from miscible to immiscible when $\alpha$ increase
beyond $1$. No such change, however, happens at $\alpha =1$ in the
high angular momentum limit. This qualitative difference in behavior
occurs because the presence of a vortex lattice naturally modulates
the density of each component, with the high density regions of one
fluid coincident with the low density regions of the other.  Thus the
system is effectively
phase separated whenever staggered vortex lattices are present, even
for $\alpha<1$.  In particular, the vortex lattice near $\alpha=1$
(above or below) is made up of alternating rows of vortices of each
component, (see Fig. 1), and the system therefore contains stripes in
which one component has a high density and the other component has a
very low density.  As $\alpha$ increases, the stripes become more
pronounced.
%To reiterate, the vortex lattices 
%provide a mechanism for
%gradually turning on phase separation, and this process begins far
%below $\alpha=1$.

{\bf Final Remarks:} The diversity of the vortex lattice structures in
the two-component Bose gas has once again demonstrated the rich
properties of these systems. Our calculation, based on exact
evaluation of the vortex energy, assumes a perfect lattice.
Considering the long relaxation times in clouds of dilute atoms, one
might see more complicated structures, where patches of vortex domains
are separated by defects or grain boundaries.
Nevertheless, the underlying equilibrium structure should be reflected
within each vortex domain.

So far, we have only discussed two-components systems with simple
interpenetrating Bravais lattices.  Our method is more general in that
it allows the exact evaluation of the energy of an arbitrary regular
lattice (with however complicated unit cell decoration).  Such
structures may be favored when the particles in each component have
different numbers, trapping potentials, or masses (as in the case of
$^{23}$Na-$^{87}$Rb mixtures).

{\bf Appendix: } The general form of a vortex lattice in the lowest
Landau level is $\Psi(x,y) = f(w)e^{-r^2/2d^2}$, where $w=x+iy$ and
$f$ is an entire function whose zeros form a regular lattice $\{ b =
nb_{1} + n_{2}b_{2} \}$, where $n_{i}$ are integers, and $b_{1}$ and
$b_{2}$ are the complex basis vectors, ($b_{2}=b_{1}(u+iv)$). Since
the Jacobi theta function $\theta( (x+iy)/b_{1}, u+iv)$ is an
entire function with exactly these zeros, we have $f(w) = \theta(
\zeta, \tau) h(\zeta)$, where $\zeta = (x+iy)/b_1 = \bar x+i\bar y$, 
$\tau=u+iv=b_{2}/b_{1}$, and $h(\zeta)$ is an
entire function without zeros.  To ensure the normalizability of
$\Psi$, this function can only be of the form $h(\zeta) = {\rm
  exp}(c_{1}\zeta + c_{2}\zeta^2)$.
It is straightforward to show that
\begin{eqnarray}
& |\theta(\zeta, \tau)|^2 = \sum_{m}(-1)^{m} e^{2\pi i m \bar x} 
 e^{-\pi v m^2/2} L_{m}&
\label{thetasqure} \\
&L_{m} =\displaystyle \frac{1}{2} \sum_{m'}
\left( 1 - e^{i\pi
 (m+m')}
 \right) e^{\left(i\pi u m - 2\pi \bar y
-\pi vm'/2\right)m^\prime} &
\label{Lm} 
\end{eqnarray}
By applying the Poisson summation formula to to Eq. (\ref{Lm}), we
have
\begin{equation}
L_{m} = \sqrt{\frac{2}{v}} \sum_{k} (-1)^{(m+1)k} 
e^{\left(-\pi(k + um + 2i\bar y)^2/2 v \right)}. 
\label{sumLm} \end{equation}
We thus have
\begin{equation}
|\theta(\zeta, \tau)|^2 = \left[\frac{1}{v_{c}} \sum_{\bf K} g_{\bf K}
e^{i{\bf K}\cdot {\bf r}} \right]
e^{2\pi y^2/v_{c}}
\label{finalthetat2} \end{equation}
where ${\bf r} = x \hat x + y \hat y$,
${\bf K} = (2\pi m \hat x -
2\pi (n+um)/v \hat y)/b_1$, and $g_{\bf K}$ is given by (\ref{gK2}).
The density of the system is then $|\Psi({\bf r})|^2 = |\theta(\zeta,
\tau)|^2 |e^{c_{1}\zeta + c_{2}\zeta^{2}}|^2 e^{-r^2/d^2}$. For a
vortex lattice with inversion symmetry about the origin ${\bf r}=0$,
we have $c_{1}=0$. In addition, if the cloud's envelope is
cylindrically symmetric, we have $c_2=\pi/(2 v_c)$, which gives Eqs.
(\ref{new}),
(\ref{sigma}),
(\ref{f})
and
(\ref{key}).
%As they are both analytic functions sharing the same zeros,
%the product in Eq. (\ref{sigmatheta}) 
%and our expression $f(w) =
%\theta(\zeta, \tau)e^{\pi w^2/2v_{c}}$ must be identical 
%up to a multiplicative function of the form
%is that the product in
%Eq. (\ref{sigmatheta}), 
%is ``proportional" to 
%the Weierstrass $\sigma$-function, 
%which is in
%turn ``proportional" to 
%and the Jacobi $\theta$ function are all identical up to an entire
%function of the form  
%$e^{c_{1} \zeta + c_{2} \zeta^2}$.
%
%.  (See for
%example, Tkachenko's study of vortex lattices in liquid helium
%\cite{tkachenko}.)  By ``proportional", we mean up to a entire function
%of the form $e^{c_{1} \zeta + c_{2} \zeta^2}$.
Similar approaches have been used by Tkachenko \cite{tkachenko} 
and Abrikosov \cite{abrikosov} in their
respective studies of $^4$He and type two superconductors.

This work is supported by NASA Grants NAG8-1441, NAG8-1765, and by NSF
Grants DMR-0109255, DMR-0071630.  

%\vspace{0.2in}

\begin{figure}
  \centerline{\includegraphics*[width=0.75\columnwidth]{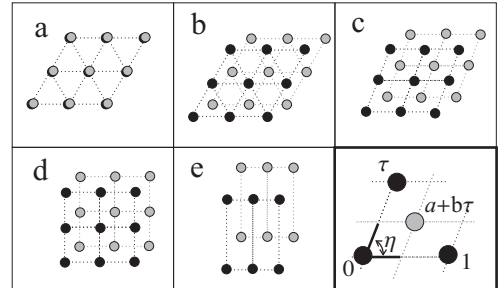}}
\caption{Phases of the two-component lattice:  black and grey dots
  represent vortices of each of the two fluids.  The panels a) through
  e) show the vortex structure in each of the phases described in the
  text.  The final panel depicts the geometry of the lattices; the
  black and grey dots respectively occupy positions in the complex
  plane $\{m+n \tau\}$, and $\{(a+m)+(b+n)\tau\}$, where $m,n$ are
  integers.  All minimal-energy configurations have $a=b$.
}\label{phasediagram}
\end{figure}
\begin{figure}

  \centerline{\includegraphics*[width=0.95\columnwidth]{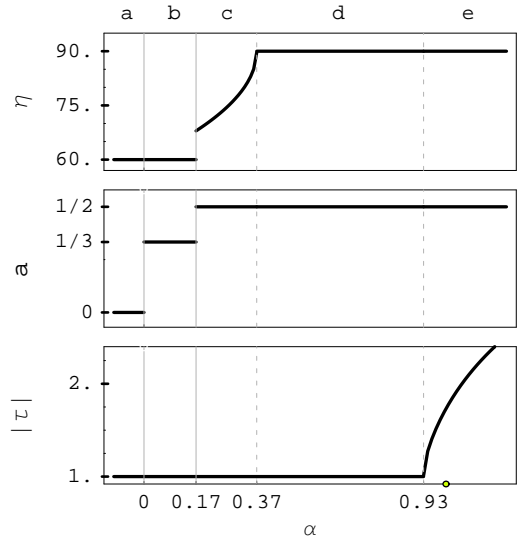}}
\caption{The parameters of the vortex lattice as a function of
  $\alpha=g_{12}/\sqrt{g_1g_2}$, a measure of the importance of
  interactions between unlike atoms.  The phases, labeled a through e
  are illustrated in Fig.~\ref{phasediagram} along with the parameters
  $\tau=|\tau|e^{i\eta}$ and $a$.  Solid and dashed vertical lines
  respectively denote first and second order phase transitions.  The
  open circle on the horizontal axis indicates $\alpha=1$.
}\label{cp}
\end{figure}

\end{document}